\documentclass[fleqn,usenatbib]{mnras}

\usepackage[T1]{fontenc}
\DeclareRobustCommand{\VAN}[3]{#2}
\let\VANthebibliography\thebibliography
\def\thebibliography{\DeclareRobustCommand{\VAN}[3]{##3}\VANthebibliography}

\usepackage{graphicx}	% Including figure files
\usepackage{amsmath}	% Advanced maths commands
\usepackage{amssymb}	% Extra maths symbols
\usepackage[normalem]{ulem} % strike through text

\def\gtorder{\mathrel{\raise.3ex\hbox{$>$}\mkern-14mu
             \lower0.6ex\hbox{$\sim$}}}
\def\ltorder{\mathrel{\raise.3ex\hbox{$<$}\mkern-14mu
             \lower0.6ex\hbox{$\sim$}}}

\title[Neutron Star Mergers]{Photometric prioritization of neutron star merger candidates}

\author[Ofek et al.]{Eran O. Ofek$^{1}$\thanks{E-mail: eran.ofek@weizmann.ac.il},
Nora L. Strotjohann$^{1}$,
Iair Arcavi$^2$, 
Avishay Gal-Yam$^{1}$, 
Doron Kushnir$^{1}$,
\newauthor
Eli Waxman$^{1}$,
Mansi M. Kasliwal$^{3}$,
Andrew Drake$^{3}$,
Matthew Graham$^{3}$,
Josiah Purdum$^{4}$,
\newauthor
Ben Rusholme$^{5}$, %\author[0000-0002-9998-6732]{
Yashvi Sharma$^{6}$,
Roger Smith$^{4}$,
Avery Wold$^{6}$,
Brian F. Healy$^{7}$
\\
$^{1}$Department of Particle Physics and Astrophysics, Weizmann Institute of Science, 76100 Rehovot, Israel.\\
$^2$School of Physics and Astronomy, Tel Aviv University, Tel Aviv 69978, Israel.\\
$^3$Division of Physics, Mathematics, and Astronomy, California Institute of Technology, Pasadena, CA 91125, USA.\\
$^4$Caltech Optical Observatories, California Institute of Technology, Pasadena, CA 91125, USA.\\
$^5$ IPAC, California Institute of Technology, 1200 E. California Blvd, Pasadena, CA 91125, USA.\\
$^6$Cahill Center for Astronomy and Astrophysics, California Institute of Technology, Pasadena, CA 91125, USA.\\
$^7$School of Physics and Astronomy, University of Minnesota, Minneapolis, MN 55455, USA.
}

\pubyear{2021}

\begin{document}

\label{firstpage}
\pagerange{\pageref{firstpage}--\pageref{lastpage}}
\maketitle

% Abstract of the paper
\begin{abstract}

Rapid identification of the optical counterparts of Neutron Star (NS) merger events discovered by gravitational wave detectors may require observing a large error region and sifting through a large number of transients to identify the object of interest. Given the expense of spectroscopic observations, a question arises: How can we utilize photometric observations for candidate prioritization, and what kinds of photometric observations are needed to achieve this goal?
NS merger kilonova exhibits low ejecta mass ($\sim5\times10^{-2}$\,M$_{\odot}$) and a rapidly evolving photospheric radius (with a velocity $\sim0.2c$). As a consequence, these sources display rapid optical-flux evolution. Indeed, selection based on fast flux variations is commonly used for young supernovae and NS mergers. In this study, we leverage the best currently available flux-limited transient survey—the Zwicky Transient Facility Bright Transient Survey—to extend and quantify this approach.
We focus on selecting transients detected in a 3-day cadence survey and observed at a one-day cadence. We explore their distribution in the phase space defined by $g-r$, $\dot{g}$, and $\dot{r}$. Our analysis demonstrates that for a significant portion of the time during the first week, the kilonova AT\,2017gfo stands out in this phase space. It is important to note that this investigation is subject to various biases and challenges; nevertheless, it suggests that certain photometric observations can be leveraged to identify transients with the highest probability of being fast-evolving events.
We also find that a large fraction ($\approx75$\%) of the transient candidates
with $\vert\dot{g}\vert>0.7$\,mag\,day$^{-1}$, are cataclysmic variables or active galactic nuclei with radio counterparts.

\end{abstract}


\section{Introduction}
\label{sec:intro}

The classification of astronomical transients usually requires spectroscopic resources. To date, only about 10\% of the transients reported worldwide to the Transient Name Server (TNS\footnote{\url{https://www.wis-tns.org/}}) have spectroscopic observations (\citealt{Kulkarni2020arXiv_FollowupOpticalTransients}).
The alternative of photometric classification has partial success (\citealt{Poznanski+2002PASP_Supernova_Color_Classification}).

A related topic is
how to prioritize follow-up resources for candidates of optical (ultraviolet to infrared) emission that originate from compact-object mergers and additional types of fast-evolving transients.
With the increase in the gravitational-wave detector horizon, deeper searches are required, which in turn will increase the number of transient candidates that are found in the
gravitational-wave error region.
Furthermore, in order to study the kilonova spectral evolution from early times after the merger, very rapid classification or at least target prioritization is required.

A feature of NS-merger-like transients is that they involve ejecta with:
low-mass
($M_{\rm ej}\sim5\times10^{-2}$\,M$_{\odot}$),
opacity of $\ltorder1$\,cm$^{2}$\,g$^{-1}$,
and high velocities
($v_{\rm ej}\sim0.2$c),
where $c$ is the speed of light (e.g. \citealt{Rosswog+2018A&A_GW170817_Analysis, Waxman+2018_GW170817_LC, Nakar2020_EM_CompactBinaryMergers_NS}).
In the first few days after the merger,
the ejecta are optically thick,
resulting in a very fast evolution
of the light curve (compared to supernovae and most other transients).
For example, in the case of AT\,2017gfo, the optical counterpart of GW\,170817 (\citealt{Abbott+2017_GW170817_MultiMessengerObservations}), the ejecta velocity can be estimated from the photometric observations alone (see e.g., \citealt{Waxman+2018_GW170817_LC}).
Specifically, when enough observations are available,
and the distance is known,
it is possible to fit the photospheric radius as a function of time and obtain its velocity.
Here, we examine a simpler approach of using the color and flux derivatives of the transients.
Using flux derivatives is not new.
For example, \cite{Khazov+2016_FlashSearchSample} identified young core-collapse SNe based on their rise-time evolution,
while \cite{Andreoni+2021ApJ_FastTransientsSearches_ZTF} applied this for NS-merger candidates.
\cite{Bianco+2019PASP_PrestoColor_FastTransientIdentification} solved the reverse problem and used this to estimate the minimal cadence requirements for an LSST-like survey to identify fast transients
like compact-object merger afterglows,
and rapidly evolving blue transients
(e.g., \citealt{Drout+2014_Rapidly_Evolving};
\citealt{Ho+2021_ZTF_RapidlyEvolvingTransients}; \citealt{Ho+2019_AT2018cow_Radio};
\citealt{Ho+2019ApJ_AT2018gep_RapidlyEvolvingTransient};
\citealt{Ofek+2010_PTF09uj_windBreakout};
\citealt{Ofek+2021_AT2018lqh_RapidlyEvolving}).
The \cite{Bianco+2019PASP_PrestoColor_FastTransientIdentification} estimate is based on simulated Type-Ia SNe, and a few observed core-collapse SNe.
Here, we inspect the position
of various transients,
including AT\,2017gfo, in the
color vs. magnitude time-derivatives phase space.
We find that a simple combination
of the transient color
and the magnitude-derivatives in
one or two bands is
sufficient, in most cases, for removing a large fraction of unrelated transient sources
and efficiently prioritizing
follow-up observations.
%We provide a table of the observed probability density function of transients in the $g-r$, $\dot{g}$, and $\dot{r}$ (i.e., magnitude's time derivatives) phase space.

In \S\ref{sec:SN}, we present the sample of background transients that we use,
while in \S\ref{sec:ColorDeriv}, we calculate the color and magnitude-derivatives of objects
in this sample, as well as for AT\,2017gfo.
In \S\ref{sec:Dist}
we present the distribution of transients in the $g-r$, $\dot{g}$, and $\dot{r}$ phase space, in \S\ref{sec:Outliers} we discuss the nature of the common outliers in this phase space, and
we conclude in \S\ref{sec:Disc}.
Throughout the paper, the dot symbol above the band name 
denotes a time derivative.
Throughout the paper, we use 
tools from \cite{Ofek2014_MAAT}.

\section{Transients sample}
\label{sec:SN}

To estimate the background of transients we use a sample of sources,
which are part of the Zwicky Transient Facility
(ZTF; \citealt{Bellm+2019_ZTF_Overview}, \citealt{Graham+2019_ZTF_objectives}, \citealt{Dekany+2020_ZTF_Camera}, \citealt{Bellm+2019_ZTF_Scheduler})
Bright Transient Survey
(BTS; \citealt{Fremling+2020ApJ_ZTF_BTS_BrightTransientSurvey_Spectroscopy, Perley+2020ApJ_ZTF_BTS_BrighTransientSurvey_StatSample})
found during 2018--2021\footnote{BTS detected some known transients that were discovered and named before the beginning of the survey.}. Here, we summarize how these transients are identified in near real time.

The ZTF pipeline (\citealt{Masci+2019_ZTF_Pipeline}) does the image calibration and produces difference images based on the \cite{Zackay+2016_ZOGY_ImageSubtraction} algorithm.
PSF-fit photometry runs on the difference images and sources in different epochs are matched to generate light curves. The pipeline generates about $10^6$ alerts per night (\citealt{Patterson+2019PASP_ZTF_Alert_Stream}, \citealt{Masci+2019_ZTF_Pipeline}). These are then filtered down to a small number of candidates ($\sim500$ per night in 2018 \citep{Fremling+2020ApJ_ZTF_BTS_BrightTransientSurvey_Spectroscopy}
and $<50$ per night after
that \citep{Perley+2020ApJ_ZTF_BTS_BrighTransientSurvey_StatSample}).
At the end of the night an astronomer reviews this list, judges which of the candidates are likely genuine astrophysical transients, and triggers spectroscopy based on the expected luminosity at peak (\citealt{Fremling+2020ApJ_ZTF_BTS_BrightTransientSurvey_Spectroscopy}). Candidates with prior variability are discarded, such that a large fraction of active galactic nuclei (AGNs) and cataclysmic variables (CVs) are rejected. However, some are saved as transients nevertheless.
Transient candidates are usually classified with the SEDMachine on the P60 telescope (\citealt{Ben-Ami+2012_SEDM}; \citealt{Blagorodnova+2018_SEDM}).

We downloaded alert photometry through the Fritz broker \citep{Duev+2019_RealBogousZTF, Kasliwal+2019_GROWTH_Marshal}. However, nightly alert packages are also publicly available from the ZTF Alert Archive\footnote{\url{https://ztf.uw.edu/alerts/public/}} for nights since June 2018. The ZTF Avro Alert package\footnote{\url{https://github.com/ZwickyTransientFacility/ztf-avro-alert}} provides the tools to read the alerts and filter them by object ID to extract only the ones that were selected for the BTS\footnote{\url{https://sites.astro.caltech.edu/ztf/bts/explorer.php}} sample. Alternatively, alert light curves can be obtained from different community brokers. % such as MARS, Lasair, ANTARES, and ALeRCE. 

The BTS is based on the public three-day cadence survey, and only these observations are used to identify transients. However, many
of these transients have ZTF observations obtained at a higher cadence.
Here we use all the available ZTF data
(i.e. not only the BTS survey data).
As discussed in the next section, we are using only observations which have one day (or faster) cadence.
This may introduce biases when trying to estimate the probability distribution function of fast-evolving transients in
the color and magnitude-derivatives phase space.
For example, fast transients have a higher probability
of being missed.
Nevertheless, given the large number of transients
in the BTS sample, including some fast-evolving transients, it is likely
good enough for our objective of target prioritization
and reducing the follow-up load.

BTS aims to obtain spectra for all bright transients and is 97\% complete for SNe brighter than $18$th magnitude, 93\% for $18.5$ and 75\% for $19$ \citep{Perley+2020ApJ_ZTF_BTS_BrighTransientSurvey_StatSample}.

The fact that the BTS sample magnitude limit
is about two magnitudes brighter compared
to the ZTF limiting magnitude is important to our purpose.
Specifically, without this magnitude limit,
our sample will be dominated by transients detected near maximum light,
and will be biased towards objects with slow magnitude evolution (i.e., the magnitude derivative near the peak is minimal).

We corrected the light curves for the Galactic extinction
(\citealt{Schlegel+1998_DustMapsReddening}),
assuming $R_{V}=3.08$ (\citealt{Cardelli+1989_Extinction}).
We only used photometric measurements that have errors smaller than $0.3$\,mag.
We note that using different measurement errors cut of $0.05$\,mag or $0.15$\,mag do~not changes the results significantly.
We also removed sources brighter than magnitude 14.5.
The final sample we use is an order of magnitude smaller than the
full BTS sample. The main reason for this is our requirement for selecting
objects that were observed with a 1-day cadence and small photometric errors.

The color and magnitude-derivative distribution of sources
on the sky depends on when the last search of the sky was conducted.
For example, a survey with one month cadence will mostly detect old transients
that evolve slowly, while the new transients found by a one-day cadence survey
will mostly be young transients or transients near their maximum light.
For that reason, we present results for two samples.
The first sample, is for all the observations, regardless of the age of the transient,
and the second sample, is for only the first epoch of each unique transient.
In Table~\ref{tab:Type} we list the number of magnitude-derivative measurements and unique objects for each transient class. 
\begin{table}
	\centering
	\caption{Number of BTS transients and photometric measurement by class.}
	\label{tab:Type}
	\begin{tabular}{lcc} 
	    \hline
	    Type & Measurements & Unique objects \\
	           &  & \\
	    \hline
                  ? &    113 &    33\\ 
                 AGN &    711 &    80\\ 
                AGN? &     46 &     8\\ 
                  CV &    180 &    44\\ 
                 CV? &      8 &     4\\ 
                 LBV &     16 &     3\\ 
               Other &      1 &     1\\ 
              SLSN-I &     19 &     4\\ 
             SLSN-II &     17 &     3\\ 
               SN II &    471 &    91\\ 
              SN IIP &     36 &     8\\ 
              SN IIb &     11 &     3\\ 
              SN IIn &    205 &    30\\ 
               SN Ia &   1180 &   315\\ 
           SN Ia-91T &     40 &     8\\ 
          SN Ia-91bg &      3 &     1\\ 
           SN Ia-CSM &      9 &     1\\ 
           SN Ia-pec &     15 &     2\\ 
              SN Iax &      7 &     2\\ 
               SN Ib &     33 &    10\\ 
             SN Ib/c &      8 &     2\\ 
               SN Ic &     50 &    12\\ 
            SN Ic-BL &     18 &     4\\ 
                 TDE &     27 &     4\\ 
                Nova &      1 &     1\\

    \hline
sum & 9657 & 966\\
	    \hline
	\end{tabular}
\end{table}

\section{Calculation of the color and magnitude-time derivatives}

\label{sec:ColorDeriv}

For each transient in our sample, we calculate its $g$- and $r$-band magnitude time derivatives ($\dot{g}$, and $\dot{r}$, respectively) and
its $g-r$ color at various epochs.
In order to calculate the derivatives and colors
we select light curves with at least three observations in two consecutive nights
in the $g$ and $r$ band, respectively. For each filter, at least one of the three observations of each filter is in one night,
and at least two others are in the previous or next night.
Practically, this was done by searching for time windows of at least 18 hours and less than 30 hours (i.e., two consecutive nights) in which there are at least three $g$-band observations,
and independently also,
at least three $r$-band observations that were taken on two successive nights.
The ZTF data is almost always collected with at least two observations per night.
Therefore, our requirement for three data points within the time window is not diluting the sample size significantly.
We selected only photometric data points with an uncertainty smaller than 0.3\,magnitude.
For each set of data points fulfilling these criteria, we fitted a first-degree polynomial as a function of time (\citealt{Press+2002_Book_NumericalRecepies, Gould2003_LinearLeastSquares_Chi2}), separately to the $g$ and $r$ band data. The first-degree polynomial was fitted after subtracting the window mid-time from the time of the observations\footnote{Failing to subtract the midpoint will introduce a strong covariance between the fitted slope and intersection (e.g., \citealt{Gould2003_LinearLeastSquares_Chi2}).}. The slope of the polynomial gives us $\dot{g}$ or $\dot{r}$, while the intersection gives us the mean magnitude, from which the color can be determined\footnote{For convenience, we call it magnitude derivative ($\dot{g}$), but in practice, this is a numerical derivative estimated over 1\,day time scale.}.
We also rescaled the ZTF $r$-band errors by a factor of $\cong1.17\cong\sqrt(1.38)$.
The reason for this is that when doing the linear fit for each time window,
without the scaling, the average, over all the fits, of the $\chi^{2}/dof$, was about $0.96$ and $1.38$, for the $g$, and $r$-band, respectively.

%The requirement for at least three observations in each band allows us to estimate the uncertainty in the color and magnitude-derivative measurements independently of the reported errors. The errors were estimated by scaling the formal errors\footnote{Adding a cosmic error term and/or scaling the formal errors is discussed in \cite{Yee+2012ApJ_MOA-2011-BLG-293Lb_Chi2Norm}.} such that the $\chi^{2}$ per degree of freedom will be one\footnote{Implemented using MATLAB {\tt lscov} function.} (e.g., \citealt{Yee+2012ApJ_MOA-2011-BLG-293Lb_Chi2Norm}). We added in quadrature, to all the errors, the estimated photometric accuracy calibration of ZTF, of about $0.015$\,mag (\citealt{Masci+2019_ZTF_Pipeline}).

%
Finally, these magnitudes are corrected for Galactic extinction.
We note that our $\dot{g}$ were measured over one day time scale,
and that for fast-evolving transients the magnitude derivative over shorter time scales may be larger.
Table~\ref{tab:BTS} provides all the derived values and their estimated errors.
In total, we have 3225 
epochs
for 674
unique transients.
The transient types include:
active galactic nuclei (AGN),
cataclysmic variables (CV),
luminous blue variables (LBV),
superluminous SN (SLSN),
Type II, IIn, Ia, Iax, Ia-91T, Ia-CSM, Ib, Ic, Ic broad line,
and tidal disruption event (TDE) candidates (see Table~\ref{tab:Type}).
The classifications are derived from the original ZTF-BTS classifications (\citealt{Fremling+2020ApJ_ZTF_BTS_BrightTransientSurvey_Spectroscopy, Perley+2020ApJ_ZTF_BTS_BrighTransientSurvey_StatSample}), which were made through human inspection of the spectra and are therefore subjective.

\section{Distribution of the color and magnitude-time derivatives}
\label{sec:Dist}

% For the prioritization/classification process, we assume that the observer has access to at least three or four data points taken in two bands (preferably $g$ and $r$ bands), and can calculate $g-r$, $\dot{g}$, and optionally $\dot{r}$. Additional bands are discussed later in this section.

\begin{figure*}
\centerline{\includegraphics[width=15cm]{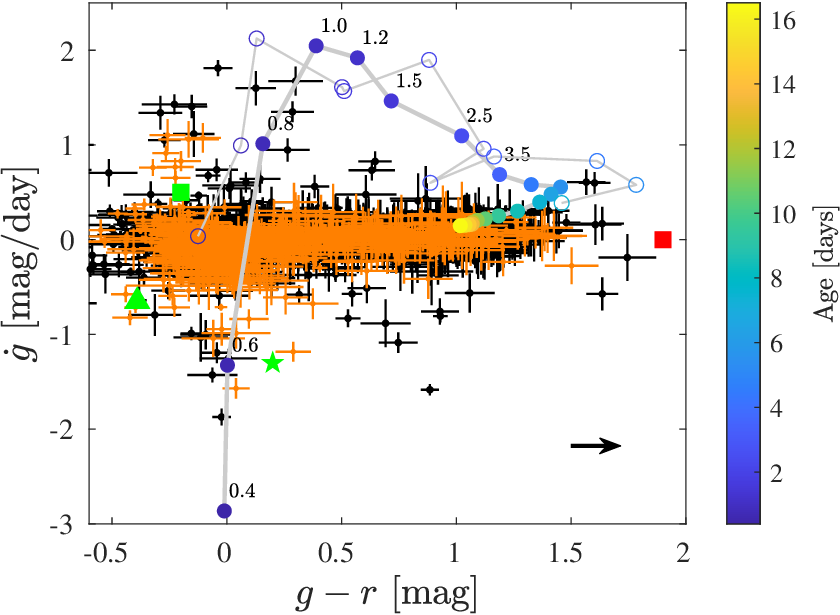}}
\caption{The $g-r$ color vs. the time-derivative of the $g$-magnitude for the BTS transient sample and AT\,2017gfo, the optical afterglow of GW\,170817.
The thick gray line shows the observed time evolution of AT\,2017gfo (\citealt{Waxman+2018_GW170817_LC}), where the color of the filled circles, as well as the numbers, indicates the kilonova age since the NS merger (see colorbar).
The thin gray line and the empty circles are the same but for the \citealt{Arcavi+2017ApJ_AT2017gfo_Observations} compilation.
The black dots show transients from the BTS sample (\citealt{Fremling+2020ApJ_ZTF_BTS_BrightTransientSurvey_Spectroscopy}, \citealt{Perley+2020ApJ_ZTF_BTS_BrighTransientSurvey_StatSample}).
The orange dots are for the first epoch of each unique transient.
The green square shows AT\,2018lqh during the decay, immediately after maximum light (\citealt{Ofek+2021_AT2018lqh_RapidlyEvolving});
the red box represents the M85-OT\,1 luminous red nova transient during the plateau phase (\citealt{Kulkarni+2007Natur_M85OT1_LRN}, \citealt{Ofek+2008ApJ_M85OT1_Environment});
the green triangle shows SN\,2019hgp (\citealt{Gal-Yam+2022Natur_AT2019hgp_Icn_SN});
and the green star shows the early observations of
SN\,2013fs taken during the light curve rise (\citealt{Yaron+2017_PTF13dqy_HighIonization_FlashSPectroscopy}).
The black arrow shows the direction of the reddening with $E_{B-V}=0.2$\,mag (\citealt{Cardelli+1989_Extinction}).
All magnitudes were corrected for Galactic extinction.
\label{fig:gr_gdot_SN_GW170817}}
\end{figure*}

\begin{figure}
\centerline{\includegraphics[width=7.5cm]{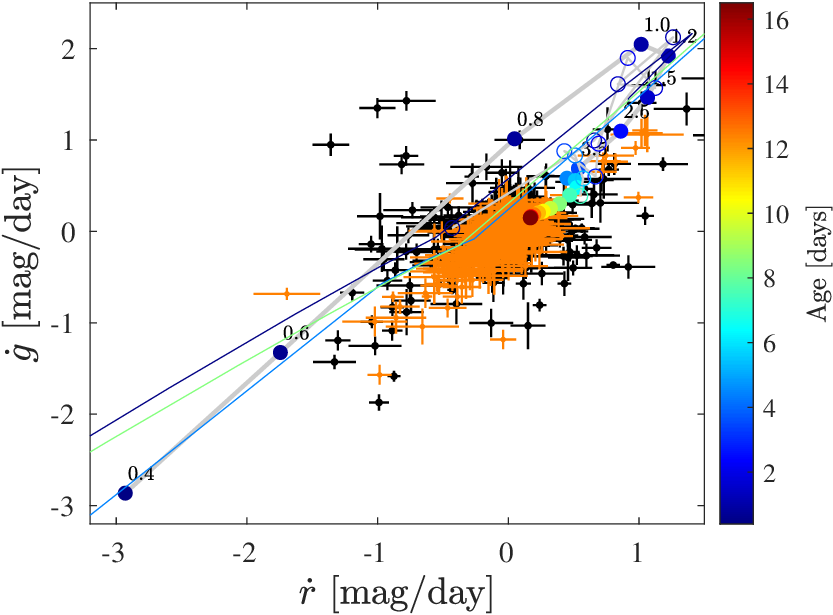}}
\caption{$\dot{r}$ vs. $\dot{g}$. Symbols are like in Figure~\ref{fig:gr_gdot_SN_GW170817}.
\label{fig:rdot_gdot_SN_GW170817}}
\end{figure}

Figure~\ref{fig:gr_gdot_SN_GW170817} shows the $g-r$ vs. $\dot{g}$ values of all BTS transients that pass our cuts, as well as the position as a function of time of AT\,2017gfo. Figure~\ref{fig:rdot_gdot_SN_GW170817} presents the $\dot{r}$ vs. $\dot{g}$ for the BTS sample and AT\,2017gfo.
In figures~\ref{fig:gr_gdot_SN_GW170817}--\ref{fig:rdot_gdot_SN_GW170817}, black points are for the full sample (all epochs), while orange points show the first epoch for each unique transient.
We added to the plots a few transients that are not in the BTS sample.
This includes fast transients
(so-called Fast Blue Optical Transients; FBOT) like AT\,2018lqh (\citealt{Ofek+2021_AT2018lqh_RapidlyEvolving}),
Luminous red novae
(M85\,OT-1; \citealt{Kulkarni+2007Natur_M85OT1_LRN}),
and early detected SNe:
SN\,2019hgp (\citealt{Gal-Yam+2022Natur_AT2019hgp_Icn_SN}), and SN\,2013fs (\citealt{Yaron+2017_PTF13dqy_HighIonization_FlashSPectroscopy}).
In order to estimate the uncertainty in
the color and magnitude-derivative of
AT\,2017gfo,
we present two compilations of photometric binned data (see \citealt{Abbott+2017_GW170817_MultiMessengerObservations}),
from \cite{Waxman+2018_GW170817_LC}
and \cite{Arcavi+2018ApJ_GW170817_ImportanceEarlyObservations}.
In both cases, the data was collected from
\cite{Andreoni+2017PASA_AT2017gfo_Observations},
\cite{Arcavi+2017ApJ_AT2017gfo_Observations},
\cite{Coulter+2017Sci_AT2017gfo_Observations},
\cite{Cowperthwaite+2017ApJ_AT2017gfo_Observations},
\cite{Diaz+2017ApJ_AT2017gfo_Observations},
\cite{Drout+2017Sci_AT2017gfo_Observations},
\cite{Evans+2017Sci_AT2017gfo_Observations_NuSTAR},
\cite{Hu+2017SciBu_AT2017gfo_Observations},
\cite{Kasliwal+2017Sci_AT2017gfo_Observations},
\cite{Lipunov+2017ApJ_AT2017gfo_Observations},
\cite{Pian+2017Nature_GW170817_Spectra_AT2017gfo_Observations},
\cite{Pozanenko+2018ApJ_AT20127gfo_Observations},
\cite{Shappee+2017Sci_AT2017gfo_Observations},
\cite{Smartt+2017Natur_AT2017gfo_Observations},
\cite{Tanvir+2017ApJ_AT2017gfo_Observations},
\cite{Troja+2017Natur_AT2017gfo_Observations},
\cite{Utsumi+2017PASJ_AT2017gfo_Observations}, and
\cite{Valenti+2017ApJ_AT2017gfo_Observations}.
The \cite{Waxman+2018_GW170817_LC} compilation
estimates the magnitude in logarithmicly-spaced time bins\footnote{The bins in which the magnitude-derivatives were calculated are equal to the time between the following bin and the previous bin.}
by fitting a polynomial to all the measurements within each time bin,
while \cite{Arcavi+2018ApJ_GW170817_ImportanceEarlyObservations} binned the data in 0.1\,day bins.
Table~\ref{tab:AT2017gfo} lists the $g-r$, $\dot{g}$, and $\dot{r}$
derived from the \cite{Waxman+2018_GW170817_LC} 
compilation of AT\,2017gfo observations (see \citealt{Abbott+2017_GW170817_MultiMessengerObservations}).
For completeness, we also provide in this table the color and magnitude derivative calculated in some other selected bands.
\begin{table*}
	\centering
	\caption{BTS transients, magnitudes and magnitude derivatives after Galactic extinction correction.
$\delta$ symbols indicate uncertainties
 The first five lines are displayed. MJD refers to the mid-time of the g-band observations. $\Delta{T}_{\rm g}$ is the time since the first $g$-band detection.
For repeating events
this refers to the time since the first detected event. The full table is available in the electronic version of this manuscript.}
	\label{tab:BTS}
	\begin{tabular}{lllrrrrrrrrr} 
	    \hline
	    Name & Type & MJD & $g-r$ & $\delta{(g-r)}$ & $\dot{g}$ & $\delta{\dot{g}}$ & $\dot{r}$ & $\delta{\dot{r}}$ & $\Delta{T}_{\rm g}$ & $\chi^{2}_{\rm g}$/dof & $\chi^{2}_{\rm r}$/dof\\
	     &  & day & mag & mag & mag/day & mag/day & mag/day & mag/day & day & & \\
	    \hline
           AT2022fmn   &      AGN&  59676.732&  $0.13$   &0.08&  $-0.276$& 0.110&  $-0.484$& 0.116& 1418.5 &   1.5/3&    0.0/2\\
           AT2016blu   &      LBV&  58875.953&  $ -0.36$ &0.06&  $ 0.371$& 0.061&  $ 0.996$& 0.112&  603.7 &   0.5/2&    1.8/2\\
           AT2016blu   &      LBV&  58898.832&  $ -0.27$ &0.07&  $ 1.052$& 0.076&  $ 1.548$& 0.134&  626.6 &   7.6/5&    1.2/4\\
           AT2016blu   &      LBV&  58903.670&  $  1.74$ &0.13&  $-0.188$& 0.246&  $-0.088$& 0.055&  631.4 &   0.1/1&    0.5/8\\
           AT2016blu   &      LBV&  58940.884&  $ -0.69$ &0.10&  $ 0.736$& 0.074&  $ 1.183$& 0.191&  668.7 &  32.8/4&    1.1/2\\

	    \hline
	\end{tabular}
\end{table*}

\begin{table*}
	\centering
	\caption{AT\,2017gfo colors and magnitude-derivatives as a function of time.}
	\label{tab:AT2017gfo}
	\begin{tabular}{rrrrrrrrrrrrr} 
	    \hline
	     t & $g-r$ & $\dot{g}$ & $\dot{r}$ &  $\dot{u}$ & $\dot{i}$ & $\dot{J}$ & $\dot{H}$ & $\dot{K}$ & $u-g$ & $r-i$ & $J-H$ & $H-K$\\
	     day& mag & mag/day & mag/day & mag/day & mag/day & mag/day & mag/day & mag/day & mag & mag & mag & mag \\
	    \hline
	    $ 0.4$ &  $-0.01$ & $-2.86$ & $-2.93$ &   $-2.18$ & $-3.01$ & $-2.58$ & $-3.04$ & $-3.43$ & $ 0.05$ & $-0.13$ & $-0.54$ & $-0.56$ \\ 
$ 0.6$ &  $ 0.00$ & $-1.32$ & $-1.74$ &   $-0.64$ & $-1.96$ & $-1.92$ & $-2.37$ & $-2.70$ & $ 0.19$ & $-0.11$ & $-0.45$ & $-0.48$ \\ 
$ 0.8$ &  $ 0.16$ & $ 1.01$ & $ 0.05$ &   $ 1.49$ & $-0.40$ & $-1.13$ & $-1.49$ & $-1.62$ & $ 0.33$ & $-0.04$ & $-0.36$ & $-0.43$ \\ 
$ 1.0$ &  $ 0.39$ & $ 2.05$ & $ 1.02$ &   $ 2.62$ & $ 0.50$ & $-0.55$ & $-0.98$ & $-1.19$ & $ 0.38$ & $ 0.07$ & $-0.30$ & $-0.44$ \\ 
$ 1.2$ &  $ 0.57$ & $ 1.92$ & $ 1.22$ &   $ 2.66$ & $ 0.80$ & $ 0.01$ & $-0.38$ & $-0.67$ & $ 0.56$ & $ 0.17$ & $-0.19$ & $-0.34$ \\ 
$ 1.5$ &  $ 0.72$ & $ 1.46$ & $ 1.07$ &   $ 2.05$ & $ 0.76$ & $ 0.23$ & $ 0.06$ & $-0.04$ & $ 0.74$ & $ 0.27$ & $-0.13$ & $-0.30$ \\ 
$ 2.5$ &  $ 1.02$ & $ 1.10$ & $ 0.86$ &   $ 1.47$ & $ 0.68$ & $ 0.33$ & $ 0.21$ & $ 0.14$ & $ 1.29$ & $ 0.53$ & $ 0.01$ & $-0.23$ \\ 
$ 3.5$ &  $ 1.19$ & $ 0.69$ & $ 0.54$ &   $ 0.83$ & $ 0.43$ & $ 0.28$ & $ 0.16$ & $ 0.09$ & $ 1.49$ & $ 0.64$ & $ 0.11$ & $-0.17$ \\ 
$ 4.5$ &  $ 1.33$ & $ 0.58$ & $ 0.45$ &   $ 0.62$ & $ 0.34$ & $ 0.25$ & $ 0.12$ & $ 0.02$ & $ 1.58$ & $ 0.75$ & $ 0.24$ & $-0.08$ \\ 
$ 5.5$ &  $ 1.45$ & $ 0.56$ & $ 0.51$ &   $ 0.54$ & $ 0.45$ & $ 0.28$ & $ 0.15$ & $ 0.04$ & $ 1.56$ & $ 0.86$ & $ 0.37$ & $ 0.03$ \\ 
$ 6.5$ &  $ 1.41$ & $ 0.48$ & $ 0.53$ &   $ 0.45$ & $ 0.53$ & $ 0.33$ & $ 0.20$ & $ 0.09$ & $ 1.54$ & $ 0.87$ & $ 0.51$ & $ 0.13$ \\ 
$ 7.5$ &  $ 1.36$ & $ 0.40$ & $ 0.47$ &   $ 0.36$ & $ 0.52$ & $ 0.39$ & $ 0.27$ & $ 0.15$ & $ 1.50$ & $ 0.85$ & $ 0.63$ & $ 0.25$ \\ 
$ 8.5$ &  $ 1.27$ & $ 0.30$ & $ 0.39$ &   $ 0.27$ & $ 0.46$ & $ 0.47$ & $ 0.35$ & $ 0.23$ & $ 1.46$ & $ 0.76$ & $ 0.76$ & $ 0.36$ \\ 
$ 9.5$ &  $ 1.18$ & $ 0.25$ & $ 0.32$ &   $ 0.24$ & $ 0.39$ & $ 0.49$ & $ 0.39$ & $ 0.27$ & $ 1.44$ & $ 0.71$ & $ 0.89$ & $ 0.48$ \\ 
$10.5$ &  $ 1.13$ & $ 0.22$ & $ 0.27$ &   $ 0.21$ & $ 0.33$ & $ 0.47$ & $ 0.39$ & $ 0.29$ & $ 1.43$ & $ 0.63$ & $ 0.97$ & $ 0.59$ \\ 
$11.5$ &  $ 1.09$ & $ 0.20$ & $ 0.23$ &   $ 0.20$ & $ 0.27$ & $ 0.46$ & $ 0.35$ & $ 0.29$ & $ 1.42$ & $ 0.57$ & $ 1.06$ & $ 0.68$ \\ 
$12.5$ &  $ 1.07$ & $ 0.18$ & $ 0.19$ &   $ 0.18$ & $ 0.22$ & $ 0.42$ & $ 0.32$ & $ 0.28$ & $ 1.41$ & $ 0.56$ & $ 1.19$ & $ 0.73$ \\ 
$13.5$ &  $ 1.07$ & $ 0.16$ & $ 0.17$ &   $ 0.17$ & $ 0.19$ & $ 0.35$ & $ 0.29$ & $ 0.28$ & $ 1.42$ & $ 0.52$ & $ 1.27$ & $ 0.76$ \\ 
$14.5$ &  $ 1.06$ & $ 0.16$ & $ 0.17$ &   $ 0.16$ & $ 0.17$ & $ 0.32$ & $ 0.25$ & $ 0.25$ & $ 1.43$ & $ 0.52$ & $ 1.31$ & $ 0.74$ \\ 
$15.5$ &  $ 1.04$ & $ 0.15$ & $ 0.17$ &   $ 0.15$ & $ 0.17$ & $ 0.36$ & $ 0.21$ & $ 0.22$ & $ 1.43$ & $ 0.52$ & $ 1.42$ & $ 0.74$ \\ 
$16.5$ &  $ 1.02$ & $ 0.15$ & $ 0.17$ &   $ 0.15$ & $ 0.17$ & $ 0.38$ & $ 0.18$ & $ 0.22$ & $ 1.43$ & $ 0.51$ & $ 1.61$ & $ 0.71$ \\ 
		\hline
	\end{tabular}
\end{table*}

There are some 
differences between
the two photometric compilations
of AT\,2017gfo, especially at early times $\ltorder1$\,day.
This is presumably because
it is difficult to estimate the magnitude-time-derivative when the magnitude differences between
adjacent observations are small.
The \cite{Waxman+2018_GW170817_LC} compilation seems to result
in smoother estimates of $g-r$, $\dot{g}$, and $\dot{r}$.
Therefore, for our objective, they are likely preferred.
Regardless of the actual magnitude-derivative of AT\,2017gfo,
we expect the $\dot{g}$ and $\dot{r}$ to be negative
and have larger values than other transients at some early time after the merger ($\ltorder0.5$\,day). 
To summarize,
AT\,2017gfo seems, as expected, to evolve
faster than other transients,
for the majority of the time.
The two compilations demonstrate that accurate estimates of $\dot{g}$, $\dot{r}$, and $g-r$ are not straightforward to obtain.
Nevertheless, even with the current uncertainties, it is possible to separate compact-object mergers from the majority of transients.

\begin{figure}
\centerline{\includegraphics[width=8cm]{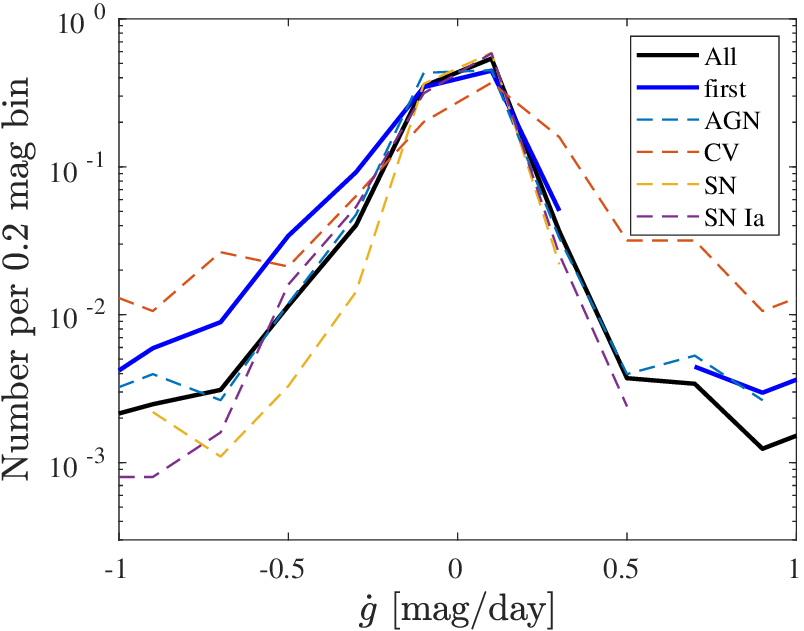}}
\caption{$\dot{g}$ distribution (histograms) for selected classes of transients. The black line represents the entire sample and the blue line is the first epoch of all unique transients.
The histograms are normalized such their sum is unity.
The heavy dashed lines shows the $\dot{g}$ distribution calculated using a magnitude-error cutoff of $0.1$\,mag instead of $0.3$\,mag.
\label{fig:gdot_hist}}
\end{figure}

Figure~\ref{fig:gdot_hist} presents the $\dot{g}$ distribution
for several classes of objects,
including all measurements (heavy black),
first measurement for each unique object (heavy blue),
and selected classes of transients.
The first important feature, seen in this plot, is that the first observation
of each transient tends to have a more extreme $\dot{g}$ value.
Second is that some classes of objects, most noticeably CVs, tend to produce a higher fraction
of extreme $\dot{g}$ values.

\begin{figure}
\centerline{\includegraphics[width=7.5cm]{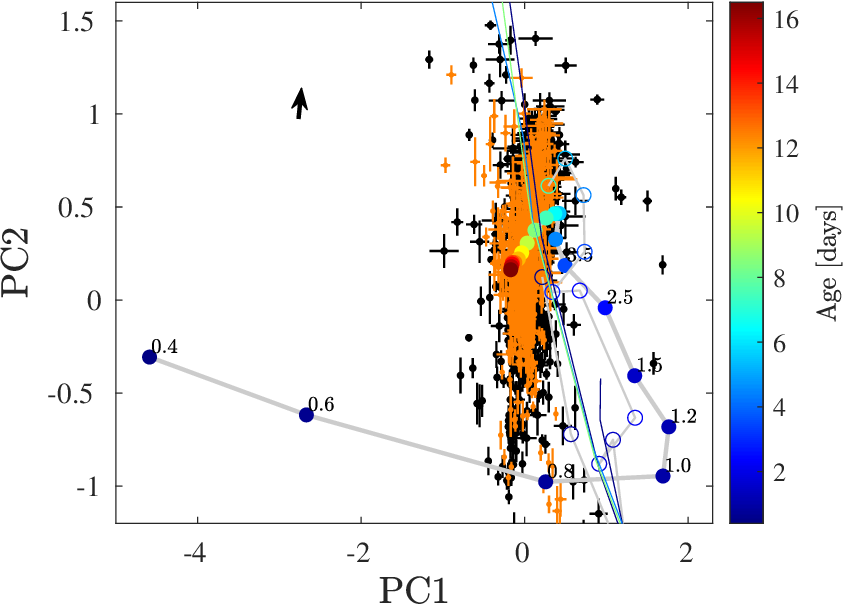}}
\caption{Projection of $g-r$, $\dot{g}$ and $\dot{r}$ into a plane (PC1 vs. PC2; Equation~\ref{eq:PCA}) in which the AT\,2017gfo observations reside (see text for details). Symbols are like in Figure~\ref{fig:gr_gdot_SN_GW170817}.
\label{fig:PCA13_rdot_gdot_SN_GW170817}}
\end{figure}

To visually demonstrate that it is possible to separate AT\,2017gfo-like events from more common transients,
we would like to find a transformation
of the three observables $g-r$, $\dot{r}$ and $\dot{g}$
that separates the two populations well and along the entire time evolution of AT\,2017gfo.
For simplicity, we use a linear transformation
and perform a principal-components
analysis on the $g-r$, $\dot{r}$ and $\dot{g}$ of all the AT\,2017gfo observations from the \cite{Waxman+2018_GW170817_LC} compilation.
We find that 99.8\% of the variance content is
in the first and second principal components.
Figure~\ref{fig:PCA13_rdot_gdot_SN_GW170817} shows the first principal component (PC1) vs. the second principal component (PC2) for all the transients in the BTS sample as well as AT\,2017gfo.
The equations for the principal components, after the extinction correction, are:
\begin{eqnarray}
    PC1 = 0.1471(g-r) + 0.8702\dot{g} - 0.4703\dot{r},\\
    PC2 = 0.7337(g-r) - 0.4149\dot{g} - 0.5381\dot{r},\\
    PC3 = 0.6634(g-r) + 0.2659\dot{g} + 0.6995\dot{r}.
    \label{eq:PCA}
\end{eqnarray}
This is not necessarily the best linear transformation that separates the two
populations, but it seems good enough for
presentation purposes.
Figure~\ref{fig:PCA13_rdot_gdot_SN_GW170817} demonstrates that the information content in the $g-r$ color is not negligible.
Specifically, it is possible to separate between AT\,2017gfo and the BTS transient population up to day $\sim10$, after the merger.
There are multiple ways to use this information for target selection.
This includes: preferring targets that are located at low-density regions of the phase space we explore in this work, and using Table 2 to estimate the rough observed probability distribution of events in this phase space\footnote{All the tables, code, and approximate probability distribution of events in this phase space are available from: {\url https://github.com/EranOfek/NS-mergers-photometric-prioritization}.}

\section{Outliers in $\dot{g}$}
\label{sec:Outliers}

It is worthwhile to inspect the nature of the outliers seen
in Figure~\ref{fig:gr_gdot_SN_GW170817}.
In Table~\ref{tab:Outliers} we list all the 47 outliers with $\vert\dot{g}\vert>0.7$\,mag\,day$^{-1}$.
An interesting finding is that about $50$\% of the entries in this table are CVs,
and about $25$\% are AGNs.
In most cases, these can be easily identified by the fact
that they likely had a previous flare (or large derivative phase) in the data (i.e., $\Delta{T}_{\rm g}$ larger than a few days).
The 13 events associated with AGNs are due to six unique objects,
while among 24 events associated with CVs, there are 15 unique objects.

The fast rise time of dwarf novae (that likely dominates our sample) is well known. For example, \cite{Cannizzo+Mattei1998ApJ_SSCug_Outbursts_Study} found that the median magnitude derivative of SS~Cyg outbursts is about 2\,mag\,day$^{-1}$, with tail extending to about 4\,mag\,day$^{-1}$.

We also added to Table~\ref{tab:Outliers} a column indicating the total flux of the brightest VLASS (\citealt{Lacy+2020PASP_VLASS_Science_SurveyDesign}) radio source found within $3''$ of the transient position\footnote{Searched using {\tt catsHTM} (\citealt{Soumagnac+Ofek2018_catsHTM}).}.
\cite{Lyke+2020ApJS_SDSSDR16_QuasarCatalog} found that about $3\%$ of the quasars
targeted by the SDSS have a radio counterpart in the FIRST catalog (e.g., \citealt{Becker+1995_FIRST}).
This is in sharp contrast to the finding that all our AGN labeled events
with $\vert\dot{g}\vert>0.7$\,mag\,day$^{-1}$ have a radio counterpart, brighter than the FIRST detection limit.

\begin{table*}
	\centering
	\caption{Subset of Table~\ref{tab:BTS}, containing outliers with $\vert\dot{g}\vert>0.7$\,mag\,day$^{-1}$. 
$\Delta{T}_{\rm g}$ for repeating events this refers to the time since the first detected event. Also added is the total $S$-band radio flux of the VLASS radio counterparts within $3''$.}
	\label{tab:Outliers}
	\begin{tabular}{llrrrrrrrr} 
	    \hline
	    Name & Type & $g-r$ & $\delta{(g-r)}$ & $\dot{g}$ & $\delta{\dot{g}}$ & $\dot{r}$ & $\delta{\dot{r}}$ & $\Delta{T}_{\rm g}$ & $F_{\rm tot}$ \\
	           &  & mag & mag & mag/day & mag/day & mag/day & mag/day & day & mJy\\
	    \hline
           AT2018ief &         AGN & $  0.93$ & $  0.03$ & $ -0.81$ & $  0.07$ & $  0.24$ & $  0.05$ & 1136.7 &   7.7\\
           AT2018cch &         AGN & $  0.53$ & $  0.05$ & $ -0.83$ & $  0.09$ & $ -0.89$ & $  0.09$ & 1489.5 &  48.6\\
           AT2019cuy &         AGN & $  0.88$ & $  0.04$ & $ -1.58$ & $  0.06$ & $ -0.87$ & $  0.05$ &   93.3 & 115.2\\
           AT2019cuy &         AGN & $  0.93$ & $  0.08$ & $ -0.76$ & $  0.14$ & $ -0.74$ & $  0.15$ &   99.4 & 115.2\\
           AT2019cuy &         AGN & $  0.75$ & $  0.08$ & $ -1.09$ & $  0.11$ & $ -0.89$ & $  0.08$ &  142.1 & 115.2\\
           AT2020dep &         AGN & $  0.69$ & $  0.11$ & $ -0.88$ & $  0.25$ & $ -0.78$ & $  0.13$ &  732.6 & 137.4\\
           AT2018efi &         CV? & $  0.29$ & $  0.08$ & $ -1.18$ & $  0.10$ & $ -0.04$ & $  0.10$ &  771.5 & \\
           AT2018ctl &          CV & $ -0.02$ & $  0.04$ & $ -1.87$ & $  0.09$ & $ -0.99$ & $  0.08$ &   93.6 & \\
           AT2018ctl &          CV & $ -0.31$ & $  0.11$ & $ -0.79$ & $  0.22$ & $ -0.40$ & $  0.20$ &  760.7 & \\
           AT2018ctl &          CV & $ -0.11$ & $  0.12$ & $ -1.03$ & $  0.26$ & $  0.15$ & $  0.13$ &  762.6 & \\
           AT2018ctl &          CV & $ -0.09$ & $  0.11$ & $ -1.00$ & $  0.15$ & $ -0.13$ & $  0.17$ &  771.7 & \\
           AT2018fag &          CV & $  0.10$ & $  0.09$ & $ -0.84$ & $  0.10$ & $ -0.46$ & $  0.14$ &  780.4 & \\
           AT2018fsg &          CV & $ -0.23$ & $  0.07$ & $ -0.71$ & $  0.10$ & $ -0.39$ & $  0.11$ &    0.7 & \\
           AT2019cmi &         AGN & $ -0.04$ & $  0.06$ & $ -1.19$ & $  0.11$ & $ -1.30$ & $  0.10$ & 1051.7 & 210.1\\
          SN2020acbm &       SN II & $ -0.40$ & $  0.05$ & $ -0.72$ & $  0.06$ & $ -0.88$ & $  0.08$ &    0.4 & \\
            SN2021dn &       SN Ia & $ -0.02$ & $  0.07$ & $ -0.94$ & $  0.17$ & $ -0.86$ & $  0.23$ &    0.8 & \\
           AT2021gem &          CV & $  0.01$ & $  0.12$ & $ -1.25$ & $  0.10$ & $ -1.02$ & $  0.20$ &   15.5 & \\
           AT2021gem &          CV & $ -0.06$ & $  0.11$ & $ -1.43$ & $  0.08$ & $ -1.33$ & $  0.16$ &   17.5 & \\
           SN2021hiz &       SN Ia & $ -0.03$ & $  0.03$ & $ -0.71$ & $  0.04$ & $ -0.62$ & $  0.04$ &    1.8 & \\
           AT2021hoz &          CV & $  0.04$ & $  0.06$ & $ -1.57$ & $  0.11$ & $ -0.98$ & $  0.09$ &    0.7 & \\
           AT2021hoz &          CV & $ -0.15$ & $  0.05$ & $ -0.99$ & $  0.07$ & $ -1.04$ & $  0.09$ &    3.8 & \\
           SN2021ont &       SN II & $ -0.42$ & $  0.08$ & $ -0.82$ & $  0.08$ & $ -0.83$ & $  0.15$ &    0.6 & \\
           SN2021too &    SN Ic-BL & $  0.04$ & $  0.15$ & $ -0.99$ & $  0.16$ & $ -1.02$ & $  0.25$ &    0.6 & \\
           SN2021vpv &   SN Ia-91T & $ -0.06$ & $  0.14$ & $ -1.04$ & $  0.20$ & $ -0.66$ & $  0.28$ &    0.7 & \\
           AT2016blu &         LBV & $ -0.27$ & $  0.07$ & $  1.05$ & $  0.08$ & $  1.55$ & $  0.13$ &  626.6 & \\
           AT2016blu &         LBV & $ -0.69$ & $  0.10$ & $  0.74$ & $  0.07$ & $  1.18$ & $  0.19$ &  668.7 & \\
           AT2016blu &         LBV & $ -0.69$ & $  0.11$ & $  1.00$ & $  0.10$ & $  0.09$ & $  0.19$ & 1019.7 & \\
           AT2018eab &          CV & $ -0.32$ & $  0.06$ & $  0.76$ & $  0.09$ & $  0.82$ & $  0.09$ & 1139.4 & \\
           AT2018eab &          CV & $ -0.29$ & $  0.09$ & $  1.34$ & $  0.18$ & $  1.36$ & $  0.25$ & 1222.0 & \\
           AT2018kfz &          CV & $ -0.26$ & $  0.08$ & $  1.10$ & $  0.17$ & $  1.05$ & $  0.10$ & 1002.5 & \\
           AT2018kkz &           - & $ -0.17$ & $  0.13$ & $  1.06$ & $  0.19$ & $  1.06$ & $  0.28$ &  750.3 & \\
           AT2019cuy &         AGN & $  0.29$ & $  0.09$ & $  1.35$ & $  0.11$ & $ -1.00$ & $  0.14$ &   82.3 & 115.2\\
           AT2019cuy &         AGN & $  0.65$ & $  0.07$ & $  0.83$ & $  0.11$ & $ -0.78$ & $  0.09$ &   85.4 & 115.2\\
           AT2019cuy &         AGN & $  0.63$ & $  0.10$ & $  0.73$ & $  0.11$ & $ -0.82$ & $  0.17$ &  790.4 & 115.2\\
           AT2019cuy &         AGN & $  0.26$ & $  0.10$ & $  0.95$ & $  0.12$ & $ -1.36$ & $  0.14$ &  815.3 & 115.2\\
           AT2019cuy &         AGN & $ -0.23$ & $  0.14$ & $  1.43$ & $  0.11$ & $ -0.78$ & $  0.22$ & 1194.3 & 115.2\\
           AT2018ctl &          CV & $ -0.15$ & $  0.07$ & $  1.41$ & $  0.12$ & $  1.02$ & $  0.11$ &   85.2 & \\
           AT2018ctl &          CV & $ -0.51$ & $  0.12$ & $  0.71$ & $  0.15$ & $ -0.17$ & $  0.28$ &  397.7 & \\
           AT2017fom &          CV & $ -0.04$ & $  0.09$ & $  0.74$ & $  0.12$ & $  0.81$ & $  0.13$ &  900.3 & \\
           AT2018imu &          CV & $ -0.20$ & $  0.09$ & $  0.83$ & $  0.11$ & $  0.75$ & $  0.21$ &  850.1 & \\
           AT2019osp &          CV & $ -0.27$ & $  0.06$ & $  0.91$ & $  0.08$ & $  0.97$ & $  0.11$ &  672.4 & \\
           AT2019vzg &          CV & $ -0.22$ & $  0.08$ & $  0.77$ & $  0.09$ & $  0.73$ & $  0.13$ &  266.7 & \\
           AT2020lbe &          CV & $ -0.04$ & $  0.06$ & $  1.81$ & $  0.09$ & $  1.63$ & $  0.11$ &   15.4 & \\
           AT2020plo &          CV & $ -0.14$ & $  0.09$ & $  1.12$ & $  0.24$ & $  0.76$ & $  0.11$ &   44.2 & \\
          AT2020abdc &          CV & $ -0.10$ & $  0.08$ & $  1.07$ & $  0.16$ & $  1.02$ & $  0.16$ &   26.2 & \\
          AT2020aeva &         AGN & $  0.30$ & $  0.12$ & $  1.67$ & $  0.15$ & $  1.51$ & $  0.18$ &   39.4 & 102.4\\
           AT2021gem &          CV & $  0.13$ & $  0.09$ & $  1.60$ & $  0.18$ & $  1.07$ & $  0.13$ &   19.3 & \\
	    \hline
	\end{tabular}
\end{table*}

\section{Discussion}
\label{sec:Disc}

We analyze the distribution of transients, identified by ZTF-BTS, in the $g-r$, $\dot{g}$, and $\dot{r}$ phase space. These transients have daily observations but were selected through a 3-day cadence survey.
The primary objective of this analysis is to establish criteria for prioritizing follow-up observations of GW events.

Within this sample, we identify three key challenges:
(i) The sample's foundation rests on a particular observation strategy and selection criteria that might not be applicable to other sky survey approaches.
(ii) The magnitude derivatives are computed on a daily timescale, whereas NS merger events (and other rapidly evolving transients) benefit from measurements taken within a span of a few hours (similar to our comparison with AT,2017gfo).
(iii) Our study is built upon specific filters.

With respect to the survey cadence, we can discuss three representative scenarios for a hypothetical survey that is being used prior to the GW trigger. The first scenario is that the last epoch was $\gtrsim3$\,day prior to the GW trigger (e.g., LSST-like cadence). In this case, the distribution of transients found in such a survey will be narrower than the distribution of the orange points in figures~\ref{fig:gr_gdot_SN_GW170817}, \ref{fig:rdot_gdot_SN_GW170817}, \ref{fig:PCA13_rdot_gdot_SN_GW170817}.
The second scenario is a sky survey with a nightly cadence\footnote{Sky surveys with a daily cadence, down to a limiting magnitude of 20–21, will likely start operating in the coming few years (e.g., \citealt{Ofek+2023PASP_LAST_Overview, BenAmi+2023PASP_LAST_Science}}. In this case, the distribution of objects in the $g-r$, $\dot{g}$, $\dot{r}$ phase space may be wider than the distribution of the orange points in figures~\ref{fig:gr_gdot_SN_GW170817}, \ref{fig:rdot_gdot_SN_GW170817}, \ref{fig:PCA13_rdot_gdot_SN_GW170817}. For example, inspecting the $\dot{g}$ distribution in Figure~\ref{fig:gdot_hist}, in the worst-case scenario the density of the wings of the distribution may be up to 3 times higher compared to their current location in this plot\footnote{This upper limit is based on the fact that the main ZTF-BTS survey is based on 3\,day cadence observations.}. This is still about two orders of magnitude below the peak of the objects in Figure~\ref{fig:gdot_hist}. 
Therefore, in this case, transients with $\dot{g}\gtrsim0.5$\,mag\,day$^{-1}$ are still rare.
The third scenario is a sky survey with multiple observations per night\footnote{In principle LAST can scan the entire visible sky from a single site, down to a limiting magnitude of 20.3, every hour (\citealt{BenAmi+2023PASP_LAST_Science}.}. In this case, we are in an unexplored territory of the transient-durations phase space. However, current limits suggest that the rate of fast transients is likely lower than the rate of transients with a few days time scale (e.g., \citealt{Perley+2020ApJ_ZTF_BTS_BrighTransientSurvey_StatSample, Ho+2023ApJ_SearchFBOT_ZTF_AT2018cow_Rate}).

With respect to the second problem (i.e., magnitude derivatives based on a one-day cadence survey). Here, the problem is that we wish to make a photometric decision in a few hours of observations, rather than one day. In this case, the question is: Are the one-day-based derivative relevant for measurements on a shorter time scale? The answer is that we do not know. However, in this case, we can argue that at least\footnote{The reason that this fraction is larger than $2/3$ is that most transients are found near their peak light.}  $2/3$ of random transients that are found, by a 3-day cadence survey, will be older than 1\,day.
Therefore, in this case, one can use the distributions of objects in Figure~1,2,4, as an order of magnitude indicator. 
A related issue is how well can we measure magnitude derivatives on a shorter time scale. This of course depends on how rapid the variations are, and on the photometric precision. Currently, ground-based sky surveys are limited to about 1\% absolute photometric accuracy (e.g., \citealt{Padmanabhan+2008_SDSS_ImprovedPhotometricCalibration, Ofek+2012_photCalib, Schlafly+2012_PS1_PhotometricCalibration}). Reaching one percent photometric accuracy for a 21-st magnitude target, using $\sim0.5$\,m class telescopes is an order of magnitude more time-efficient than obtaining spectroscopy for a similar target using a 3-m class telescope.
Assuming one percent accuracy per measurement, for a target with $\dot{g}>0.5$\,mag\,day$^{-1}$, we need a time span of $\gtrsim2$\,hr.

Another limitation is the fact that our work is based on specific filters ($g$ and $r$). In the first few days after the explosion, even transients with low-ejecta mass ($\sim10^{-2}$\,M$_{\odot}$), are optically thick, and their spectrum is roughly described by a black-body curve. Therefore, the $\dot{g}$ and $\dot{r}$ can be roughly interpolated and extrapolated, but the accuracy of this process is hard to quantify. A related problem is that filters (even if they have the same name), on different telescopes are not identical, which will introduce a color term.  In principle, these color terms can be estimated in advance.

Although obtaining observations in multiple filters is not expensive, relative to spectroscopy, sometimes such observations will not be available. In this case, we note that observations with a single filter, or even color information, without temporal information can be used, with lower effectiveness, for target prioritization.

Figure~\ref{fig:gr_gdot_SN_GW170817} demonstrates that there may be a few fast-evolving transients whose $\dot{g}$ is significantly different from zero.
However, these objects are relatively rare,
and therefore, we will still achieve our goal of reducing the number of targets for follow-up.
Among the most frequent fast-evolving transients are SNe
found at very early times after the explosion (e.g., the green star in Figure~\ref{fig:gr_gdot_SN_GW170817}; \citealt{Gal-Yam+2014_SN2013cu_FlashSpectroscopy}, \citealt{Yaron+2017_PTF13dqy_HighIonization_FlashSPectroscopy}),
fast transients (e.g., \citealt{Drout+2014_Rapidly_Evolving},
\citealt{Ofek+2010_PTF09uj_windBreakout},
\citealt{Ho+2021_ZTF_RapidlyEvolvingTransients},
\citealt{Ofek+2021_AT2018lqh_RapidlyEvolving}), and likely also
GRB afterglows (e.g., \citealt{Cenko+2013_PTF11agg}).
The expected rate of SNe brighter than
magnitude\footnote{We choose this magnitude as it is close to the limiting magnitude of some of the currently running sky surveys.}
21 is about two orders of magnitude higher than the GRB rate.
This can be estimated by multiplying the volume to which we can detect a $-18$ absolute magnitude SN (the median abs. mag. of SNe; \citealt{Perley+2020ApJ_ZTF_BTS_BrighTransientSurvey_StatSample}),
by the SNe rate ($\sim10^{-4}$\,Mpc$^{-3}$\,yr$^{-1}$; Leaman et al. 2011),
and comparing it with the GRB rate multiplied by the fraction of GRBs that are visible ($\sim500$\,yr$^{-1}$; \citealt{Cenko+2009_P60_GRB_AfterglowsObservationsCatalog}).
Furthermore, relativistic transients with no GRB emission are likely not as common as GRBs
(\citealt{Ho+2022ApJ_CosmologicalFastTransients_GRB_SearchDirtyFirballs}).

Finally, an interesting finding is that about 75\% of the $\dot{g}\gtrsim0.7$\,mag\,day$^{-1}$ events in our sample are either AGNs or CVs.
Many of these objects have recurring outbursts and can be identified in variability studies (e.g., \citealt{Wozniak+2002_OGLE-II_BuldgeVariables, Drake+2014_CRTS_PeriodicVariables, Ofek+2020_ZTF_DR1_10milion_Variables}), and be rejected as kilonova candidates.

\section*{Acknowledgements}

We thank an anonymous referee for constructive comments.
E.O.O. is grateful for the support of
grants from the 
Benozio Center,
Willner Family Leadership Institute,
Ilan Gluzman (Secaucus NJ), Madame Olga Klein - Astrachan,
Minerva Foundation,
Israel Science Foundation,
BSF-NSF, Israel Ministry of Science, Weizmann-MIT,
and the Rosa and Emilio Segre Research Award.
N.L.S. is funded by the Deutsche Forschungsgemeinschaft (DFG, German Research Foundation) via the Walter Benjamin program – 461903330.

Based on observations obtained with the Samuel Oschin Telescope 48-inch and the 60-inch Telescope at the Palomar Observatory as part of the Zwicky Transient Facility project. ZTF is supported by the National Science Foundation under Grants No. AST-1440341 and AST-2034437 and a collaboration including Caltech, IPAC, the Weizmann Institute of Science, Oskar Klein Center at Stockholm University, University of Maryland, Deutsches Elektronen-Synchrotron and Humboldt University, Los Alamos National Laboratories, the TANGO Consortium of Taiwan, University of Wisconsin at Milwaukee, Trinity College Dublin, Lawrence Livermore National Laboratories, IN2P3, France, University of Warwick, University of Bochum, and Northwestern University. Operations are conducted by COO, IPAC, and UW. 

The SED Machine is based on work supported by the National Science Foundation under Grant No. 1106171.

\section*{Data Availability}

The data presented in this paper is available in the electronic tables, while the code is accessible via GitHub\footnote{\url{https://github.com/EranOfek/AstroPack}}.

\bibliographystyle{mnras}
\bibliography{papers.bib}

% Don't change these lines
\bsp	% typesetting comment
\label{lastpage}

\end{document}